\documentclass[fleqn,twoside]{article}
\usepackage{espcrc2}

\usepackage{graphicx}
\usepackage[figuresright]{rotating}

\newcommand{\AmS}{{\protect\the\textfont2
  A\kern-.1667em\lower.5ex\hbox{M}\kern-.125emS}}

\title{Deeply Virtual Compton Scattering at HERA}

\author{J. Volmer\address[MCSD]{DESY-Zeuthen, Platanenallee 6, D-15738 Zeuthen,
    Germany}}
\begin{document}

\begin{abstract}
Deeply virtual Compton scattering has recently been studied by three HERA 
experiments, H1, ZEUS and HERMES, covering a broad range of kinematic regimes. 
We present cross section measurements of the two collider experiments in the 
kinematic region 2$<$$Q^2$$<$100~GeV$^2$ and 30$<$$W$$<$140~GeV, and compare 
them to QCD-based calculations. HERMES measurements of azimuthal asymmetries and
their kinematical dependences are presented for $Q^2$$>$1~GeV$^2$ and 
2$<$$W$$<$7~GeV.
\end{abstract}

\maketitle

\section{Introduction}

The internal structure of the nucleon has long been studied with inclusive
deep-inelastic lepton scattering. From these studies, the spin-dependent and 
spin-independent nucleon structure functions and corresponding parton 
distribution functions have been extracted. Recently, the formalism of 
generalized parton distributions (GPD) \cite{JVmul}-\cite{JVji} was introduced. 
In this framework, non-forward degrees of freedom are probed in exclusive 
processes such as meson production or deeply virtual Compton scattering (DVCS), 
the electroproduction of a single real photon on the nucleon, which has been 
studied at DESY 
\cite{H101}-\cite{Ely01} and Jefferson Lab \cite{Ste01}. This opens the way to
 access the distribution of partons in the nucleon perpendicular to its momentum
\cite{RaP01}, and to access the contribution of the orbital angular momenta of 
the quarks to the nucleon spin \cite{JiPRL97}.

\section{DVCS at HERA}

\subsection{DVCS at H1 and ZEUS}
\begin{figure}[tb]
\begin{center}
\includegraphics[width=17.5pc]{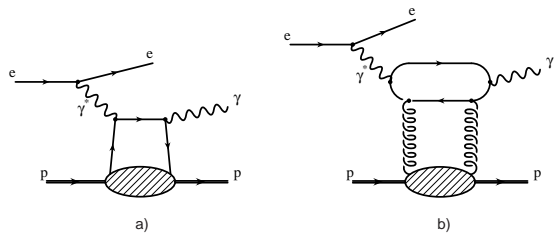}
\vspace*{-2mm}
\caption{a) leading order handbag diagram, b) NLO colour singlet 
two-gluon exchange.}
\label{fig:LO+NLO}
\end{center}
\vspace*{-5mm}
\end{figure}

The HERA storage ring at DESY provides an 820-920~GeV proton beam and a 
27.6~GeV electron or positron beam to the two collider experiments, H1 and ZEUS.
At collider energies, the leading order handbag diagram 
(Fig.~\ref{fig:LO+NLO}~a) is accompanied by a sizable contribution of colour 
singlet two-gluon exchange (Fig.~\ref{fig:LO+NLO}~b). In both cases, the final 
state is indistinguishable from that of the Bethe-Heitler (BH) process, in which
the real photon is radiated off the lepton. The BH process constitutes a 
background to the DVCS process. Since it is purely electromagnetic, its 
amplitude can be calculated exactly. This contribution is subtracted from the 
combined DVCS+BH data by means of a Monte Carlo simulation that is cross checked
with a BH-dominated sample of the experimental data
\cite{H101,Zeus01}.

\begin{figure}[htb]
\begin{center}
\includegraphics[height=5.7cm,bb=111 239 477 585, clip]{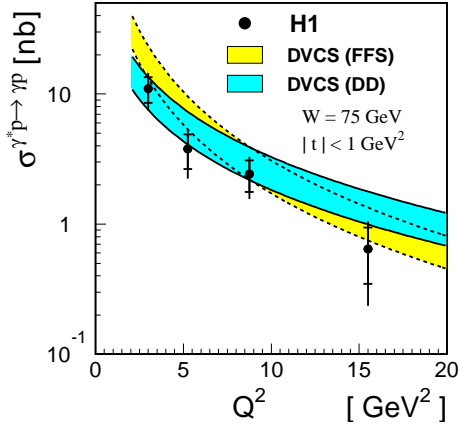}
\vspace*{-7mm}
\caption{$Q^2$ dependence of DVCS cross section for the H1 data
\protect{\cite{H101}}, GPD predictions (light shade) \protect{\cite{FFS98}}, 
a colour dipole model (dark shade) \protect{\cite{DD01}}.}
\label{fig:H1Q2}
\end{center}
\vspace*{-9mm}
\end{figure}
\begin{figure}[htb]
\begin{center}
\includegraphics[height=5.7cm,bb=111 239 477 585, clip]{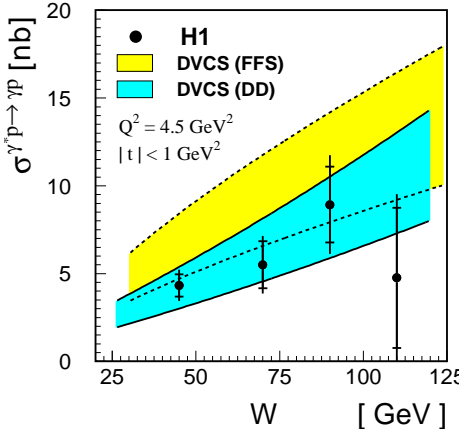}
\vspace*{-9mm}
\caption{$W$ dependence of DVCS cross section for the 1997 H1 data
\protect{\cite{H101}}, overlays as in Fig.~\protect{\ref{fig:H1Q2}}.}
\label{fig:H1W}
\end{center}
\vspace*{-8mm}
\end{figure}
\begin{figure}[htb]
\begin{center}
\includegraphics[height=5.7cm,bb=29 41 523 539,clip]{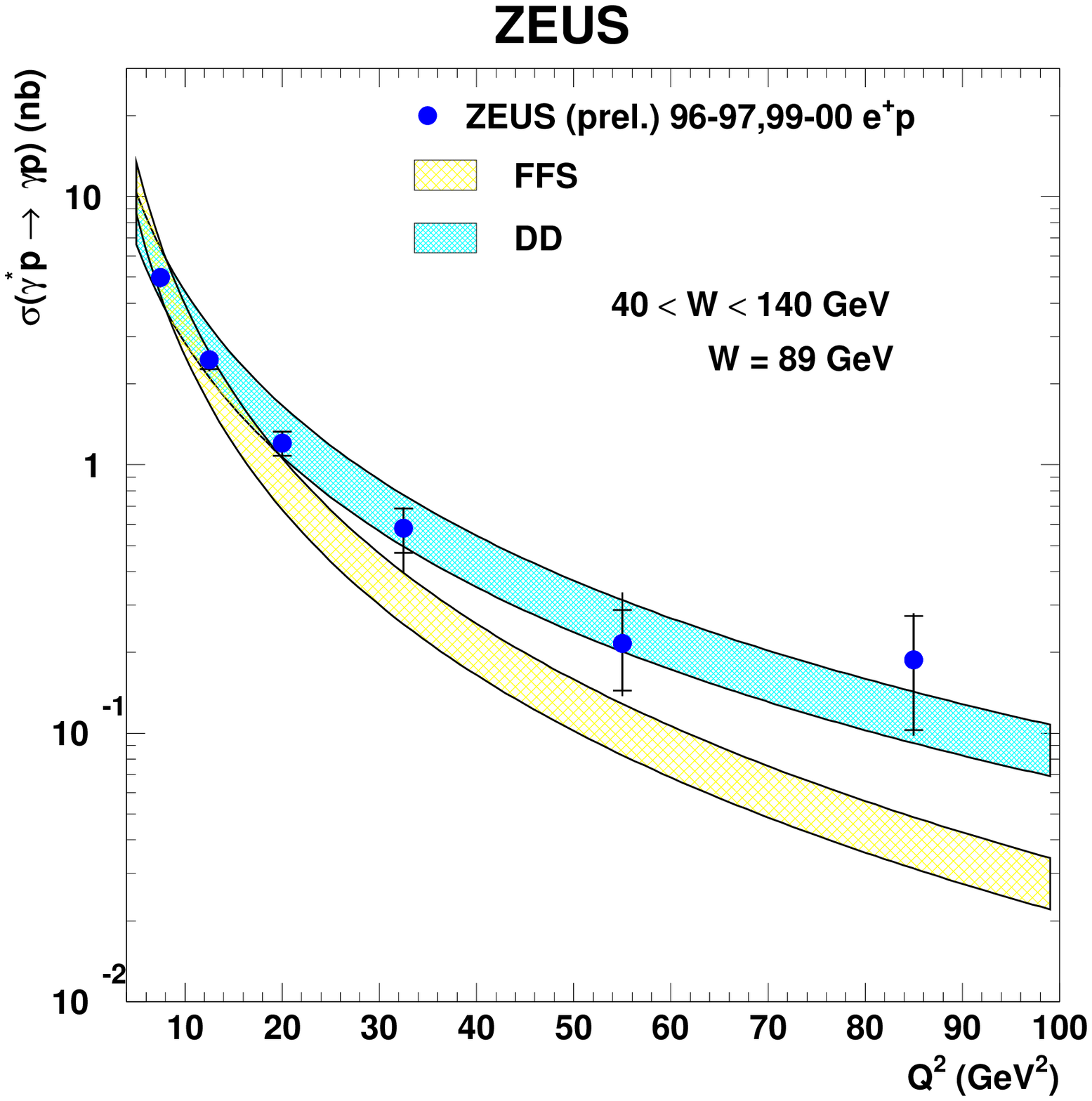}
\vspace*{-6.5mm}
\caption{$Q^2$ dependence of DVCS cross section for the ZEUS positron 
data, overlays as in Fig.~\protect{\ref{fig:H1Q2}}.}
\label{fig:ZEUS-Q2}
\end{center}
\vspace*{-8mm}
\end{figure}
\begin{figure}[htb]
\begin{center}
\includegraphics[height=5.7cm,bb=29 41 523 539,clip]{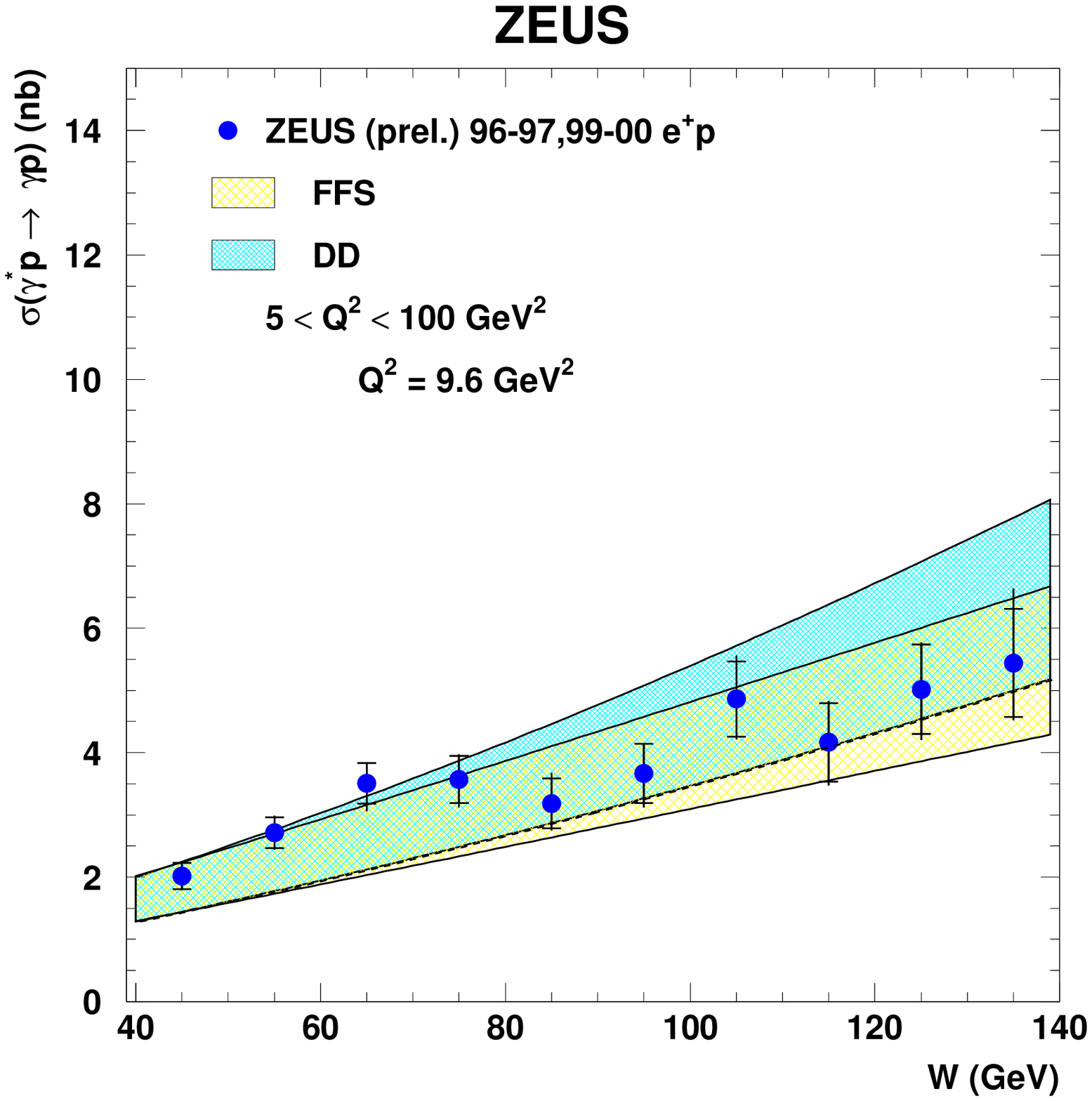}
\vspace*{-9mm}
\caption{$W$ dependence of DVCS cross section for the ZEUS positron 
data, overlays as in Fig.~\protect{\ref{fig:H1Q2}}.}
\label{fig:ZEUS-W}
\end{center}
\vspace*{-8mm}
\end{figure}
The cross section from the H1 experiment, shown in Figs.~\ref{fig:H1Q2} and 
\ref{fig:H1W}, was measured in the 1997 running period with an 820~GeV proton 
energy. The sample corresponds to an integrated luminosity of 8~pb$^{-1}$. The 
data cover a kinematic range in four-momentum transfer squared $Q^2$ from 2 to 
20~GeV$^2$, invariant mass of the photon-nucleon system $W$ from 30 to 120~GeV 
and the square of the momentum transfer to the proton $|t|$$<$1~GeV$^2$, over 
which the cross section is integrated. The data in Figs.~\ref{fig:H1Q2} and
\ref{fig:H1W} have been scaled to central values of $W$=75~GeV and 
$Q^2$=4.5~GeV$^2$, respectively. Here, also theoretical predictions for DVCS 
at H1 kinematics from a GPD-based model (FFS, \cite{FFS98}), as well as from a 
colour dipole model (DD, \cite{DD01}), are shown. In both models, the 
exponential $t$-dependence factorizes from the $Q^2$ and $W$ dependences. Since
the $t$-dependence of the cross section was not measured, upper and lower 
bounds are determined by taking values of the $t$-slope parameter $b$ of 5 
and 9~GeV$^{-2}$. A newer GPD-based approach can be found in \cite{Fre02}.

The ZEUS results are based on data taken between 1996 and 2000, and, for the 
first time, also make use of the HERA electron beam. The integrated luminosities
are 95~pb$^{-1}$ for $e^+p$ and 17~pb$^{-1}$ for $e^-p$ collisions. The data 
were analyzed in the kinematical region defined by 5$<$$Q^2$$<$100~GeV$^2$ and 
40$<$$W$$<$140~GeV, with no cut on $t$. The data were found to agree for both 
lepton charges. The $e^+ p$ data shown in Figs.~\ref{fig:ZEUS-Q2} and 
\ref{fig:ZEUS-W} are scaled to central values of $W$ = 89~GeV and $Q^2$ =
9.6~GeV$^2$, respectively. The data are overlaid with theoretical predictions 
from the same models as the H1 data. Due to the larger kinematical range and 
integrated luminosity, the ZEUS data have enough leverage and statistical 
accuracy to discriminate between different theoretical predictions and can be 
used to constrain the predicted $Q^2$ dependence of theoretical models. Here, 
the ZEUS data are seen to favour the less steep slope in $Q^2$ of the colour 
dipole model. The $W$-dependence of the data is described well in both cases, 
reflecting the steep rise in gluon density at lower $x_B$, the momentum fraction
of the struck parton in the nucleon. 

The ZEUS and H1 data cannot be compared directly due to the different kinematic 
ranges. However, if the observed functional behaviour in $Q^2$ is used to scale 
the H1 data to the ZEUS central kinematics, the two data sets agree.

\subsection{DVCS at HERMES}

\begin{figure}[tb]
\begin{center}
\includegraphics[height=6.3cm]{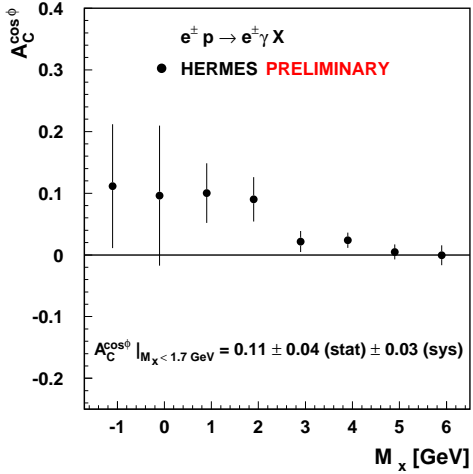}
\vspace*{-4mm}
\caption{The $\cos \phi$ moment of the beam charge asymmetry from HERMES data
from 1998 and 2000.}
\label{fig:HERMES-BCA}
\end{center}
\vspace*{-8mm}
\end{figure}

\begin{figure}[htb]
\begin{center}
\includegraphics[height=6.3cm]{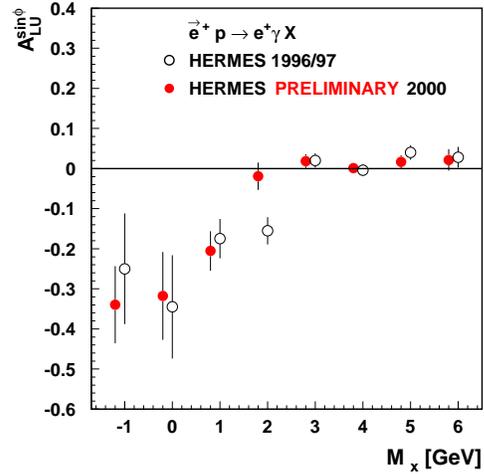}
\vspace*{-4mm}
\caption{The $\sin \phi$ moment of the beam spin asymmetry from HERMES data
from 1996/1997 (empty symbols) \protect{\cite{Ely01}} and 2000 (filled 
symbols).}
\label{fig:HERMES-BSA-96-2000}
\end{center}
\vspace*{-8mm}
\end{figure}

The fixed target experiment HERMES makes use of the 27.6~GeV polarized HERA 
positron beam incident on a storage cell that is filled with unpolarized gas
(hydrogen, deuterium, or heavier nuclei). At HERMES kinematics, the BH 
amplitude largely dominates the DVCS amplitude. However, the DVCS and BH 
amplitudes interfere, giving rise to a single-spin asymmetry, which depends on
beam charge, helicity, and the azimuthal angle $\phi$ between the lepton scattering 
plane and the reaction plane of the real photon and recoil proton.
By extracting the $\sin \phi$ and $\cos \phi$ moments of the exclusive data, 
access is obtained to certain combinations of DVCS amplitudes 
\cite{GPV01}.

The $\sin \phi$ moment of the beam spin asymmetry is calculated as
$$A^{\sin\phi}_\mathrm{LU} = \frac{2}{N} \sum_{i=1}^{N} 
\frac{\sin \phi_i}{\left( P_l \right)_i},$$
where $P_l$ is the beam polarization for each event. The $\cos \phi$ moment is 
determined similarly from the beam charge asymmetry with an unpolarized beam.

The HERMES forward spectrometer detects the scattered lepton and the real
photon, but not the recoiling proton. Therefore the exclusivity of an event is
measured through the missing mass ($M_x$) of the event, which in the exclusive
case is equal to the proton mass. Examples are shown in 
Figs.~\ref{fig:HERMES-BCA} and \ref{fig:HERMES-BSA-96-2000}, in which the $\cos 
\phi$ moment of the beam charge asymmetry for the 1998 and 2000 data, and the 
$\sin \phi$ moment of the beam spin asymmetry for the 1996/1997 data 
\cite{Ely01} and the 2000 data are shown. Asymmetries are observed at values of
$M_x$$<$1-2~GeV, in the exclusive region. At higher missing masses, where 
non-exclusive processes dominate, asymmetries vanish. The two regions overlap 
slightly, because the resolution in $M_x$ is limited by the energy resolution of
the electromagnetic calorimeter. The $\sin \phi$ measurements were confirmed at 
Jefferson Lab \cite{Ste01}. GPD-based predictions can be found in 
Refs.~\cite{Fre02,GPV01}.

\begin{figure}[tb]
\begin{center}
\includegraphics[width=\columnwidth]{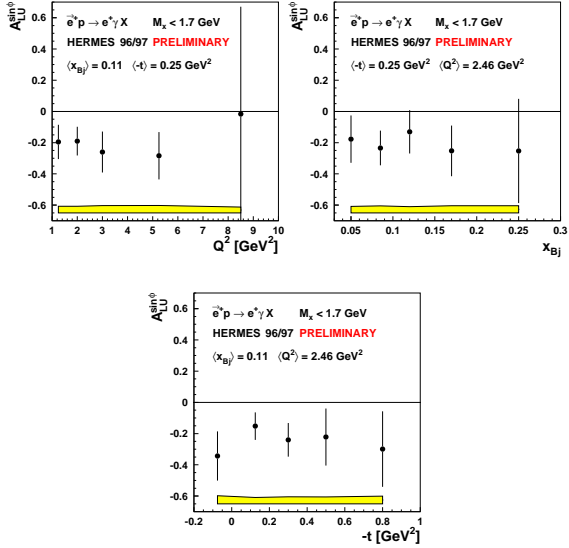}
\vspace*{-10mm}
\caption{The $Q^2$, $x_\mathrm{B}$ and $t$-dependences of the $\sin \phi$ 
moment of the beam spin asymmetry from HERMES data from 1996 and 1997.}
\label{fig:HERMES-BSA-kine}
\end{center}
\vspace*{-9mm}
\end{figure}

From the exclusive sample ($M_x$$<$1.7~GeV) of the 1996/97 data the kinematic 
dependences of the beam spin asymmetry were extracted as functions of $Q^2$, 
$x_B$ and $t$ (preliminary results shown in Fig.~\ref{fig:HERMES-BSA-kine}). In
the case of $t$, as with $M_x$, the reconstruction relies on the photon energy
measured in an electromagnetic calorimeter. Therefore some events are
reconstructed
at $-t$$<$0.

\section{Outlook}

The HERA run II starting in 2002 will be used by all three experiments to
improve the DVCS measurements. In the 2001 shutdown, spin rotators were
installed before and after the collider experiments H1 and ZEUS, thus allowing 
for spin-dependent measurements. It is planned to run alternately with 
electron and positron beams.

The H1 experiment will install a Very Forward Proton Spectrometer in 2003
\cite{VFPS00}, allowing it to measure $|t|$ at small values of $W$. In 
addition, the precise tracking data from the Backward Silicon Tracker will be 
used for a measurement of azimuthal asymmetries. 
The measurement of differences in the cross section for 
the available beam polarizations and charges will be the main goal of the ZEUS 
DVCS analysis. The HERMES experiment will improve the exclusivity of its DVCS 
sample, as well as the $t$-resolution, by installing a large acceptance recoil 
detector around the HERMES target \cite{LARD}.\\

\noindent{\bf Acknowledgements}

I would like to thank Rainer Stamen, Paul R.~Newman, Iwona Grabowska-Bo\l d and
Yuji Yamazaki for useful and enlightening discussions.

\end{document}